\documentclass[12pt]{article}
\usepackage{graphicx}
\usepackage{amssymb}
\usepackage{fullpage}

\newtheorem{theorem}{Theorem}
\newtheorem{corollary}[theorem]{Corollary}
\newtheorem{lemma}[theorem]{Lemma}

\newtheorem{definition}{Definition}

\title{Rotation Distance is Fixed-Parameter Tractable}
\author{Sean Cleary
\thanks{
	Department of Mathematics,
	City College of New York \& the Graduate Center,
	City University of New York,  New York, NY 10031,
	{\tt cleary@sci.ccny.cuny.edu}.Partial funding provided by NSF \#0811002
}
\and 
Katherine St.~John\thanks{
	Department of Mathematics \& Computer Science,
	Lehman College \& the Graduate Center,
        	City University of New York, Bronx, NY 10468,
         {\tt stjohn@lehman.cuny.edu}.  Partial funding provided by NSF  \#0513660.
}
}

\begin{document}
\maketitle

\begin{abstract}
Rotation distance between trees measures the 
number of simple operations it takes to transform one 
tree into another.  There are no known polynomial-time
algorithms for computing rotation distance.
   In the case of ordered rooted  trees,
we show that the rotation distance between two ordered
trees is fixed-parameter tractable, in the parameter, $k$,
the rotation distance.  The proof relies on the kernalization
of the initial trees to trees with size bounded by $7k$.
\end{abstract}

\section{Introduction}

\begin{figure}
\begin{center}
\includegraphics[height=2.0in]{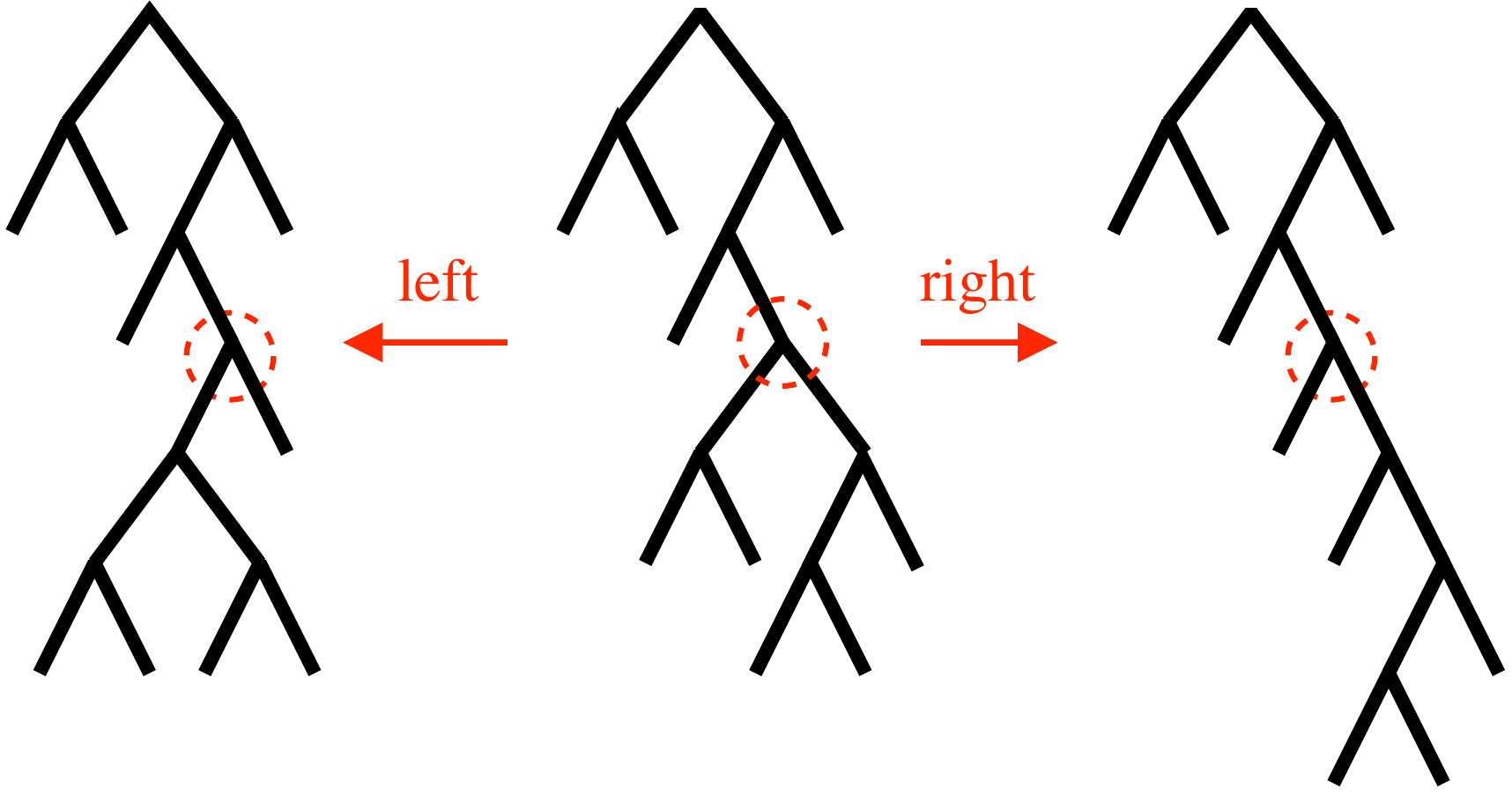}
\end{center}
\caption{\small A right
rotation at a node consists of rotating the right child of the left child of the 
node to the left child of the right child of the node.  A left rotation is defined similarly by
moving the left child of the right child of the node to the left child of the node.
The circled node in the middle tree has been rotated right to yield the
tree on the right, and similarly rotated right to yield the tree on the left.  }
\label{rot}
\end{figure}

Balancing binary search trees is an important task to ensure good performance for searches.
Rotations are simple transformations that can be used to dynamically adjust trees in an effort to keep
them reasonably balanced.  A left or right rotation at a node, shown in Figure~\ref{rot}, promotes one grandchild, demotes one child, and switches one grandchild to a different parent node.  Sequences of rotations suffice to transform a tree to any other tree with the same number of nodes. The rotation distance between two trees is the minimum number of such rotations needed to transform one to the other.
There has been a great deal of work on estimating, bounding and computing this distance.
Culik and Wood \cite{cw} gave an immediate upper bound of $2n-2$ for the distance between two trees with $n$ nodes, and in remarkable work using methods of hyperbolic volume, Sleator, Tarjan, and Thurston \cite{stt} showed not only that $2n-6$ is an upper bound, but furthermore that for all large $n$, that bound is realized.  There are no known polynomial-time algorithms for computing
rotation distance, though there are estimation algorithms of Pallo \cite{pallo}, Pallo and Baril \cite{barilpallo},
and Rogers \cite{rogers}.  The NP-hardness of this problem remains open.

We
make progress on the computational complexity of rotation distance by showing that it
is fixed-parameter tractable.  
We show the fixed-parameter tractability by reducing (or ``kernalizing'') the
common structures in the trees being compared.  
The technique of kernalizing problem instances has been very successful in 
showing the tractability of a wide range of problems \cite{Fellows2006a,Guo2007a}.
For
reduction rules, we use the ``subtree reduction'' where
identical subtrees are reduced in both, which arises in several settings
(including the study of tree pairs through Thompson's group $F$
\cite{rotbound, rotipl, rightarm}) to produce ``reduced tree pairs''  which have 
 the same rotation distance as 
the original tree pair, as shown in Figure~\ref{subtreeRuleFigure}.  This subtree
reduction rule also preserves biologically interesting
distances for phylogenetic trees (roughly, leaf-labeled,
unordered trees).  Allen and Steel \cite{allenSteel} 
showed that the biologically motivated ``tree-bisection-and-reconnection'' (TBR)
distance is preserved under the subtree reduction
rule, as well as a reduction rule that shrinks common chains,
illustrated in Figure~\ref{chainRuleFigure}.
They further show that the reduced trees have
$O(k)$  size, where $k$ is the TBR distance between the
trees.  This reduces the question of calculating the 
TBR distance to comparing trees whose size is 
dependent only on $k$, which is used to prove that TBR distance 
is fixed-parameter tractable.  Other distance calculations
have also been shown to be fixed-parameter tractable
via similar reductions \cite{uSPRfpt,bordewichSemple,Bordewich2007}.

To show that rotation distance is fixed-parameter
tractable, we follow a similar strategy to those used
for phylogenetic tree distances:  we show that the
subtree and chain reductions preserve distance
and then show that the resulting reduced trees have size 
bounded by $7k$, 
where $k$ is the rotation distance.  The
reductions can be done in linear time \cite{approx}.  
The reduced
trees can be checked in $f(k)$ time to see if their
distance is less than or equal to $k$.  While this check 
could take exponential time, it depends only
on $k$, yielding the fixed-parameter tractability of computing
rotation distance.

\section{Background}

\subsection*{Rotation Distance}

We consider ordered, leaf-labelled, rooted binary trees with $n$ interior nodes and
where each interior node has 2 children.  Such trees are
commonly called {\em extended binary trees} \cite{knuth3} or  {\em full binary trees}.
In the following, `tree' refers to such a tree with an ordering on the leaves, `node' refers
to an interior node, and `leaf' refers to a non-interior node.
Our trees will have $n$ interior nodes and  $n+1$ leaves numbered in left-to-right order
from 0 to $n$.

Right rotation at a node of a rooted binary tree is defined as as a simple change to
$T$ and is illustrated in
Figure~\ref{rot}, taking the middle tree to the right-hand one. Left rotation at a node is
the natural inverse operation.
The
{\em rotation distance}
$d_R(T_1,T_2)$ between two rooted binary trees
$T_1$ and
$T_2$ with the same number of leaves is the minimum number of rotations
needed to transform $T_1$ to $T_2$.

The specific instance of the rotation distance problem we address is:
\begin{quote}
{\sc Rotation Distance}\\
{\sc Input:} Two rooted ordered trees, $T_1$ and $T_2$ on 
$n$ leaves and parameter $k$,\\
{\sc Question:} Is the rotation distance between them, $d_R(T_1,T_2) \leq k$?
\end{quote}

There are no known polynomial time algorithms to compute rotation distance, though there are polynomial time algorithms of Rogers \cite{rogers}, Pallo\cite{pallo} and Pallo and Baril \cite{barilpallo} to estimate it.
The general difficulty of computing rotation distance comes from the lower bound.  Finding a sequence of rotations which accomplish the transformation gives only an upper bound.    

\subsection*{Edge-Flipping Distance}

Rooted binary trees correspond naturally to triangulations of polygons via a standard equivalence, described in \cite{stt}.  A basic move of changing a triangulation is to choose two adjacent triangles forming a quadrilateral in the polygon, and replace it with two triangles where the resulting quadrilateral has the opposite diagonal.  This edge-flipping move is exactly equivalent to performing a single rotation on the corresponding rooted trees.   We define {\em edge-flipping distance} between two triangulations $P_1$ and $P_2$ of an $n$-gon to be the minimal number of such edge-flipping moves to transform $P_1$ to $P_2$.  
In Lemma 3b, Sleator, Tarjan and Thurston  \cite{stt} 
show that if two subdivisions $P_1$ and $P_2$ share a common diagonal, then 
the diagonal flip distance between $P_1$ and $P_2$ is equal to the sum 
of the distances between the polygonal subdivisions obtained by the subpolygons
cut off before and after the common diagonal.   
The edge-flipping distance between $P_1$ and $P_2$ will be exactly the rotation distance of their corresponding trees.

\subsection*{Fixed-Parameter Tractability}

Roughly, the ability to efficiently calculate instances that are small, with respect
to some parameter, is called fixed-parameter tractability.  In our case,
the parameter is the distance, $k$, between the trees.  We show
that the rotation distance can be solved in time 
polynomial in the size of the input (that is, the number of leaves or nodes) 
but exponential in the size of a 
fixed-parameter, $k$ the distance.  See Downey and Fellows \cite{Downey1999}
for more details.  More formally:

\begin{definition}  A problem $\Phi$ is {\em fixed-parameter tractable} in 
a parameter $k$, if there exists some constant $c$ (independent of $k$)
and a function $f$ such $\Phi$ accepts $<\!x,k\!>$ in time $O(f(k)|x|^c)$.
\end{definition}

\section{Reduction Rules}

We use two reduction rules,  to reduce the size
of the problem.  The first rule, as mentioned above, 
replaces common subtrees with a single node.  The
second rule replaces common chains with a sequence
of 3 nodes, as done by Allen and Steel \cite{allenSteel}.
We will show in the next section 
that these reduced trees have distance bounded
by $f(k)$ for $k$ the rotation distance and $f$ 
a function that depends only on $k$, not $n$.

\begin{figure}

\begin{center}
\includegraphics[height=1.75in]{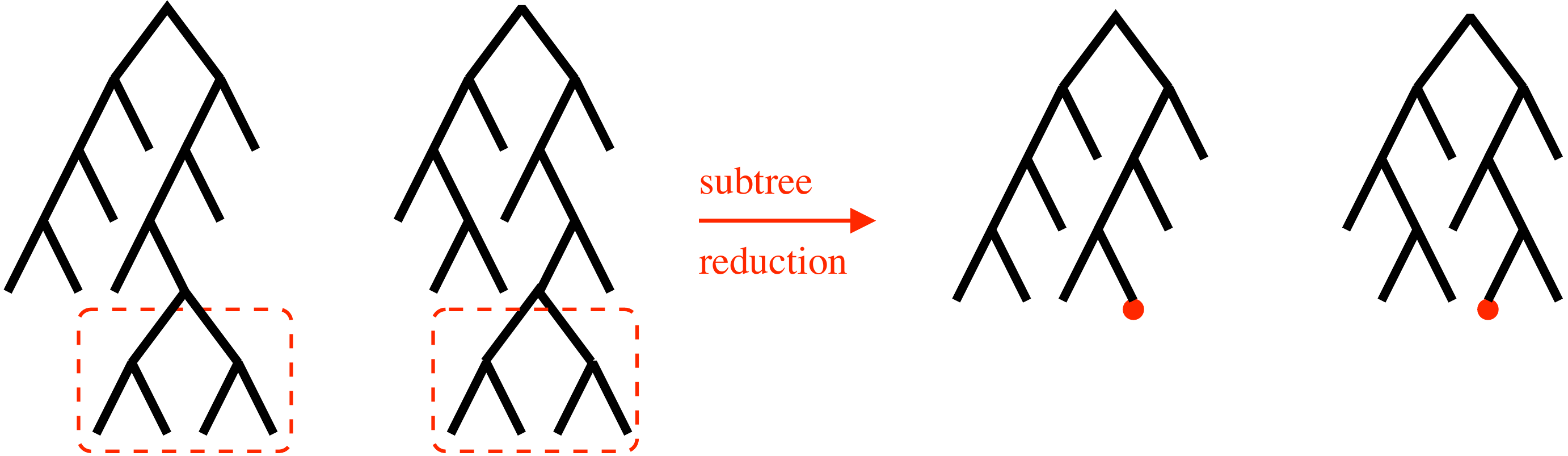}
\end{center}
\caption{\small The subtree reduction rule replaces common subtrees, in the pair of ordered
trees on the same number of leaves, 
by a placeholder of a single node.  A necessary condition is that the leaves of
subtree be labelled (numbered) identically in each of the initial trees.  Note that the subtree enclosed
by the dotted line is on the sixth through eighth leaves in the ordering in both trees.
These subtrees are replaced by a single node, indicated by a circle in the right hand
pair of trees.  The new reduced trees, like the initial trees, are on the same number
of leaves. }
\label{subtreeRuleFigure}
\end{figure}

\paragraph{Subtree Reduction:}  This straightforward reduction eliminates a
common subtree and replaces it with a single placeholder leaf.
This reduction is a natural one: Sleator, Tarjan and Thurston \cite{stt}
use it in rotation distance, interpreted there as diagonal-flip distance between triangulations of polygons.
It arises naturally in connection with the tree pair diagrams representing elements of Thompson's group $F$ \cite{rotipl}.
And it also has been used in the comparison of phylogenetic trees 
\cite{allenSteel}. 
More formally:
\begin{definition}
Given ordered trees $T_1$ and $T_2$ on $n$
leaves, if 
there exists a subtree $S$ that occurs in both $T_1$
and $T_2$, an application of the {\em subtree reduction rule} consists of replacing 
$S$ in both trees by a new leaf $s$ and 
call the new trees $T'_1$ and $T'_2$.  See 
Figure~\ref{subtreeRuleFigure}.  
\end{definition}

In the context of triangulations of polygons, common subtrees give common diagonals between the triangulations.
Lemma 3b of
\cite{stt} is that if two triangulations share a common diagonal, then any shortest part realizing the rotation distance will
not change that diagonal. Thus, in terms of the dual trees, the rotation distance is preserved 
under by the subtree reduction rule- there will be no rotations which change the common subtrees.  Thus, 
$d_{R}(T_1,T_2) = d_{R}(T'_1,T'_2)$ if $T'_1$ and $T'_2$ 
are the result of applying the subtree rule to $T_1$ and $T_2$, which gives the following lemma:

\begin{lemma}  \cite{stt}  \label{lemsubtree}
The subtree reduction rule preserves
rotation distance.
\end{lemma}

\begin{figure}
\begin{center}
\includegraphics[height=2.25in]{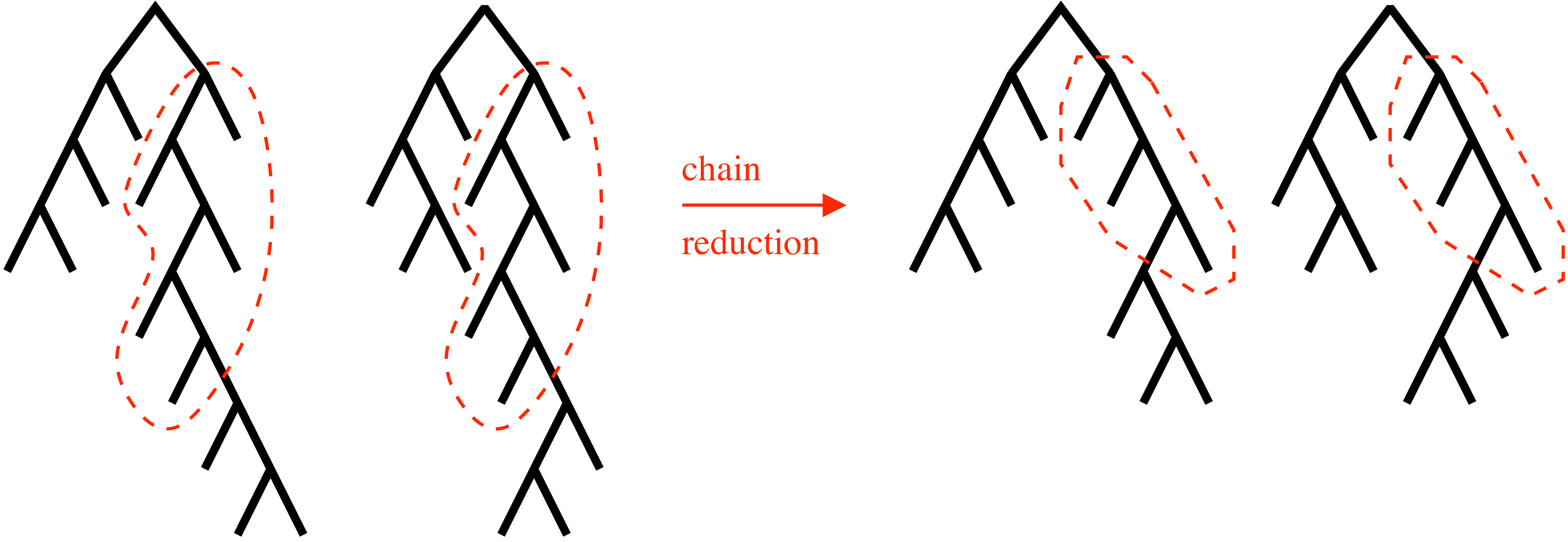}
\end{center}
\caption{\small The common chain reduction rule replaces long common chains in ordered trees on 
the same number of leaves.  Since the trees are ordered, we can assume that the leaves have
increasing labels from $0$ to $11$.  On the left, the common chain, enclosed by the dotted line, 
consists of the leaves, $11$, $4$, $10$, $5$, and $6$, traversing the chain from top to bottom.
We apply the chain reduction to this chain to yield the trees on the right.  Note that the resulting
trees are on the same number of leaves and the reduced chain contains leaves $4$, $5$, and
$9$, under the new labeling  $0$ to $9$. }
\label{chainRuleFigure}

\end{figure}

\paragraph{Chain Reduction:}  
This reduction eliminates a (potentially long) chain common to both trees and replaces it with
a short chain consisting of $3$ pendant leaves.
This is a  useful reduction rule for phylogenetic trees since 
it does not change many of the common tree distances.
Allen \& Steel \cite{allenSteel} showed that
TBR  distance is preserved when long common chains
are replaced by chains of length 3.  Variants of this
rule have also been used to show that other biologically inspired
calculations are fixed-parameter tractable \cite{uSPRfpt,bordewichSemple,Bordewich2007}.
Formally, 
\begin{definition}
Given ordered trees $T_1$ and $T_2$ on $n$
leaves, a {\em common chain}  $t_1,t_2,\ldots,t_m$ is a sequence of pendant subtrees that occur 
identically in $T_1$ and $T_2$ and for which  in both $T_1$ and $T_2$:
\begin{itemize}
\itemsep 0pt
	\item For each $j$ and $k$ such that $1\leq j < k \leq$, 
		the path between the root and the subtree $t_j$ is shorter
		than the path between the root and the subtree $t_k$.
	\item For each $j=1,\ldots,m-1$, the parent of the subtree $t_j$ is 
		the grandparent of $t_{j+1}$, and
	\item For each $j=1,\ldots,m$, $t_j$ is the left child of its parent in $T_1$
		if and only if  $t_j$ is the right child of its parent in $T_2$.
\end{itemize}
the restrictions of 
\end{definition}
\begin{definition}
Given ordered trees $T_1$ and $T_2$ on $n$
leaves, if 
there exists a common chain $t_1,t_2,\ldots,t_m$ that occurs identically in both $T_1$
and $T_2$, an application of the {\em common chain reduction rule} consists of replacing 
$t_1,t_2,\ldots,t_m$ in both trees by a chain of pendant leaves $a,b,c$, ordered the same way and 
calling the new trees $T'_1$ and $T'_2$.   
See Figure~\ref{chainRuleFigure}.
\end{definition}

We note that the chain of pendant leaves, $a,b,c$ serve as a placeholder in the 
reduction and can be oriented in any way, as long as they meet the definition of
a chain.

\section{Fixed-Parameter Tractability}

To show that rotation distance is fixed-parameter 
tractable, we first show that the common chain rule reductions
preserve rotation distance.  
This, combined with the
straightforward fact that subtree rule reductions preserve
distances (stated in \cite{stt}), gives reduced trees whose size is bounded
only by a function of $k$.   To check if the distance between
original trees is at most $k$, if suffices to check the distance
between the reduced trees. This can be
done in $O(f(k))$ time, where $f(k)$ does not 
depend on $n$, but is exponential in $k$. 

\begin{lemma} 
\label{lemchain} Let $T_1$ and $T_2$ be ordered rooted
trees, and $T'_1$ and $T'_2$ the result of applying the chain
rule reduction to these initial trees.  Then
$d_R(T'_1,T'_2) = d_R(T_1,T_2)$.
\end{lemma}

{\em Proof:}  
Let $T_1$ and $T_2$ be ordered, rooted trees with a common chain, 
$l_1,l_2,\ldots,l_m$, and let $T'_1$ and $T'_2$ the result of applying the chain
rule reduction to these initial trees.  
We consider the dual polygonal subdivisions of the trees.  In that case,
having a common chain means having a sequence of adjacent shared diagonals.
In Lemma 3b, Sleator, Tarjan and Thurston  \cite{stt} 
show that if two subdivisions $P_1$ and $P_2$ share a common diagonal, then 
the diagonal flip distance between $P_1$ and $P_2$ is equal to the sum 
of the distances between the polygonal subdivisions obtained by the subpolygons
cut off before and after the common diagonal.   
Since the sequence of common diagonals (i.e. the common chain) 
are contiguous, the distance equals the sum of the distances of the subpolygons
above and below the sequence of diagonals.  Shrinking the number
of common diagonals does not change the distances.  Thus, the 
distance between the original trees equals the distances between
the reduced trees:  $d_R(T_1,T_2) = d_R(T'_1,T'_2)$.
\hfill $\Box$

The proof of fixed-parameter tractability relies on bounding the size of 
the reduced trees, since as noted above, our algorithm
consists of reducing the trees to ones with equivalent 
distance but size bounded by a function of $k$ and then 
exhaustively trying all sequences of $\leq k$ moves
to find the distance.

\begin{theorem}
\label{lemdistsame}  Given two ordered, rooted trees, $T_1$ and $T_2$
with the same number of leaves.  Let 
$T'_1$ and $T'_2$ be the result of applying the two 
reductions rules exhaustively to the given trees. 
If $d(T_1,T_2) \leq k$, then $|T'_1| \leq 7k$.
\end{theorem}

{\em Proof:}
Given two ordered, rooted trees, $T_1$ and $T_2$
with the same number of leaves.  Assume 
that $d(T_1,T_2) \leq k$.  
We will show, by induction on $k$, that $|T'_1| \leq 7k$.

$k=1$:  Assume that $T_1$ and $T_2$ are not 
identical and that $d(T_1,T_2) = d(T'_1,T'_2) = 1$.  Then, a single
rotation, at some node $N$, transforms $T'_1$ into $T'_2$.
Since by the subtree reduction rule, identical subtrees
have been reduced, the descendants of $N$ are 
singletons.  If $N$ is not the root, the path from $N$ to
the root must be identical in both trees and can have
at most three pendant leaves (since larger common 
chains and identical subtrees have been reduced).
The other child of the root must be a singleton and 
identical in both, so, is a single leaf.   Thus, the total 
number of leaves in $T'_1$ is no more than 7.

$k>1$:  Without loss of generality, we assume that 
$k = d(T_1,T_2) = d(T'_1,T'_2)  > 1$.
Then there exists trees, $I$, with the same number of
leaves as $T_1$, and $I'$, with the same number of
leaves as $T'_1$, such that 
$d(T_1,I) = d(T'_1,I') =1$ and $d(I,T_2) = d(I',T'_2) = k-1$.
$T'_1$ and $I'$ differ by a single rotation, $r$.  

First, assume that this first rotation at node $N$ yields
a subtree identical to that in $T_2$ (that is, the subtrees 
rooted at $N$ are identical in $I$ and $T_2$).  Then, 
the trees $I'$ and $T'_2$ are not reduced (since they
have a common subtree rooted at $N$).  Since their
only difference with the tree pair $T'_1$  and $T'_2$
occurs at node $N$, a single subtree reduction will 
fully reduce $I'$ and $T'_2$.  We now argue that the 
subtree rooted at $N$ can have at most $7$ leaves.
Since the subtrees rooted at $N$ are identical in $I'$
and $T'_2$, there are no rotations below $N$ in the
tree, and $N$ has exactly $3$ leaves below it.  
As in the base case, $N$
could have a chain of at most 3 pendant leaves above
it.  If $N$ is replaced by a single node in $I'$ and $T'_2$, 
yielding new trees $I''$ and $T''_2$, then $I''$ and $T''_2$
have fewer than $|I'| -6$ leaves.  $I''$ and $T''_2$
satisfy the induction hypothesis with
$d(I',T'_2) = d(I'',T''_2) = k-1$, yielding $|I''| \leq 7(k-1)$.
Thus , $|T'_1| \leq |I''| + 6 \leq 7(k-1) + 6 \leq 7k$.

Now, assume that this first rotation at node $N$ yields
a subtree that is not identical to that in $T_2$.  Then,
the trees $I'$ and $T'_2$ are fully reduced and have
distance $k-1$, and we can
apply the induction hypothesis to them, yielding 
$|I'| \leq 7(k-1)$.  Since $|I'| = |T'_1|$, we have
$|T'_1| = 7(k-1) < 7k$.
\hfill $\Box$

From the kernalization of the problem, the fixed-parameter
tractability easily follows:

\begin{corollary} 
\label{mainthm} Rotation distance is fixed-parameter
tractable in parameter, $k$, the distance.
\end{corollary}

{\em Proof:}
Given two ordered, rooted trees, $T_1$ and $T_2$ on the
same leaf set, and a parameter $k$.  We define an algorithm
for computing ``$d_R(T_1,T_2) \leq k$'' as follows:
\begin{enumerate}
\itemsep 0pt
	\item Let $T'_1$ and $T'_2$ be the result of applying 
		the two reductions rules exhaustively to the given trees. 
	\item If $|T'_1| > 7k$, then output ``{\sc no}.''
	\item Else, $|T'_1| \leq 7k$.  Try all sequences of $k$
		or less rotations to transform $T'_1$ to $T'_2$.
		If one such sequence is successful, ouput ``{\sc yes}.''
	\item Else, if no such sequence transforms $T'_1$ to $T'_2$, 
		then output ``{\sc no}.''
\end{enumerate}

Step 1 takes time linear in the number of nodes $n$, and by Theorem~\ref{lemdistsame} the distances between the given trees and the
reduced trees are the same.  Step 3 can take no more than $(7k)^k$ steps.  Thus rotation distance is 
fixed-parameter tractable in the parameter $k$.
\hfill $\Box$

\bigskip

We note that the bound of $(7k)^k$ is crude and can be improved either by using any of the standard exhaustive methods for computing rotation distance which are exponential in $7k$, or by precomputing the rotation distances for pairs of trees of size no more than $7k$.  The Catalan numbers $C_{7k}$ count the number of trees of size $7k$, and  the number of tree pairs to consider is thus on the order of $(4^{7k})^2$.

\section{Conclusion } 

We have shown that rotation distance of ordered trees is
fixed-parameter tractable in parameter, $k$, the distance.
Step 3 of the algorithm from Section 4 is exponential in the parameter $k$, the
distance.
An obvious remaining open question is ``can rotation distance be calculated in 
polynomial time?''

\small
\bibliographystyle{plain}

 \end{document}